\documentclass[preprint,eqsecnum,aps,prd,nofootinbib]{revtex4}
\usepackage[dvips]{graphicx}
\usepackage{graphics, color}
\usepackage{amsmath}
\usepackage{enumerate}
\usepackage{mathrsfs}
\usepackage{amsfonts}

 \newcommand{\be}{\begin{equation}}
\newcommand{\ee}{\end{equation}}

\newcommand{\V}{{\mathcal V}}
\newcommand{\tV}{{\tilde{\mathcal V}}}

\newcommand{\op}{{\mathcal O}}

\usepackage{hyperref}

\begin{document}

\title{Stability in Holographic Theories with Irrelevant Deformations }

\author{Aaron J.~Amsel}
\email{aamsel@asu.edu}
\affiliation{Department of Physics and Beyond Center for Fundamental Concepts in Science, \\ Arizona State University, \\Tempe, AZ 85287}

\author{Matthew M.~Roberts}
\email{matthew.roberts@nyu.edu}
\affiliation{Center for Cosmology and Particle Physics and Department of Physics, New York University, 4 Washington Place, New York, NY 10003}

\begin{abstract}
We investigate the non-perturbative stability of asymptotically anti-de Sitter gravity coupled to tachyonic scalar fields with mass near the Breitenlohner-Freedman bound.  Such scalars are characterized by power-law radial decay near the AdS boundary, and typical boundary conditions are ``Dirichlet'' (fix the slower fall-off mode) or ``Neumann'' (fix the faster fall-off mode).  More generally though, these  ``designer gravity" theories admit a large class of boundary conditions defined by a functional relation between the two modes.  While previous stability proofs have considered boundary conditions that are deformations of the Neumann  theory, the goal of this paper is to analyze stability in designer gravity with boundary conditions that are irrelevant deformations of the Dirichlet  theory.  We obtain a lower bound on the energy using spinor charge methods and show that for the  most interesting class of such boundary conditions, the theory is always stable. We argue that the deformed theory flows to a new fixed point in the ultraviolet, which is just the Neumann theory.  We also derive a corresponding ``effective potential''  that implies stability if it has a global minimum.
\end{abstract}

\maketitle
\tableofcontents

\section{Introduction}
The bulk side of the AdS/CFT correspondence \cite{Maldacena:1997re, Witten:1998qj, Gubser:1998bc} consists of gravity coupled to various matter fields.  In particular, supergravity compactifications relevant to AdS/CFT \cite{Gunaydin:1984fk, Kim:1985ez, Gunaydin:1984wc} often contain tachyonic scalar fields with masses at or slightly above the Breitenlohner-Freedman (BF) bound \cite{BF}.  In some cases, the bulk theory can be consistently truncated so that the matter content is just scalar fields \cite{Duff:1999gh, Freedman:1999gk}.

Such theories of AdS$_{d+1}$ gravity coupled to scalar fields near the BF bound (sometimes called ``designer gravity'' \cite{HH2004}) are known to admit a large class of boundary conditions, which can be defined in terms of an arbitrary function $W$.  The scalar fields have slower fall-off than allowed by the standard asymptotically AdS boundary conditions of \cite{Henneaux:1985tv}, but nevertheless, the conserved charges have been shown to be finite and well-defined once back-reaction effects are taken into account \cite{Henneaux04,Henneaux:2006hk,Amsel:2006uf}. This paper is concerned with the conditions under which the total conserved energy is bounded from below (for other interesting applications, see e.g., \cite{Hertog:2003zs,Hertog:2003xg,HM2004,Hertog:2004gz,Hertog:2004rz,Hertog:2004jx,Hertog:2005hu,Hertog:2006rr,Craps:2007ch,Battarra:2011nt}).

The scalar field equation of motion has two linearly independent solutions, which can be represented by the coefficients of terms in the asymptotic radial expansion, labeled $\alpha$ (leading) and $\beta$ (subleading).  The standard or ``Dirichlet'' boundary condition is to turn off the slow fall-off mode, $\alpha = 0$.  More generally though, we can consider a ``deformation'' of the Dirichlet theory by taking as a boundary condition $\alpha = W'(\beta)$.  Similarly, there is a ``Neumann'' theory given by $\beta = 0$, and more generally we may choose boundary conditions of form $\beta = W'(\alpha)$.

The derivation of the energy bound proceeds by following a Witten-Nester style argument using a spinor charge  \cite{Witten:1981mf,Nester:1982tr}.
For the Dirichlet scalar boundary conditions ($\alpha = 0$), it was proven several decades ago that the energy is positive if the scalar potential is generated by a superpotential \cite{Gibbons:1983aq,Boucher:1984yx,Townsend:1984iu}.  More recently, this proof was adapted to the slow fall-off theory with boundary conditions $\beta = W'(\alpha)$ \cite{Hertog2005,Amsel:2006uf,Amsel:2007im,Faulkner:2010fh,Amsel:2011km}, where it was shown that the theory is stable if the effective potential $\V(\alpha)$ (defined below) has a global minimum \emph{and} the scalar potential admits a certain type of superpotential.   These results finally proved a conjecture about stable ground states in designer gravity that was originally given in \cite{HH2004}.

However, the analogous stability condition for deformations of the Dirichlet theory has never been given a complete treatment\footnote{Some discussion was given in \cite{Amsel:2011km} for the case where the BF bound is saturated.}.  The main goal of this paper is to fully address energy bounds in designer gravity theories with boundary conditions of the form $\alpha = W'(\beta)$.  The spinor charge calculation itself is nearly identical to that in the Neumann theory, but the implications of the resulting energy bound are somewhat surprising.  First, for any power-law boundary condition $W \sim  \beta^n$, the Dirichlet theory is always stable (assuming the appropriate superpotential exists).  This is in contrast to the Neumann case, where the existence of a lower bound on the energy tends to be heavily dependent on the choice of $W$.  Secondly, defining the appropriate effective potential $\V(\beta)$ in the Dirichlet theory is a bit subtle; it turns out not to be what one might naively expect based on the Neumann theory results.

The effective potential of \cite{HH2004} was constructed on-shell, in the sense that only the extrema of $\V(\alpha)$ gave the energy of certain static, spherically-symmetric soliton solutions.  We will verify, though, that it is possible to construct initial data for nearby solutions with the same boundary conditions for arbitrary asymptotic values of $\alpha$ and energy arbitrarily close to $\V(\alpha)$.  This implies that $\V(\alpha)$ represents a ``good'' effective potential.  A similar type of argument reveals how to define the right effective potential in the Dirichlet case.  Once again, it is this off-shell potential which appears in the energy bound and thus determines stability of the theory.

It is a general principle of AdS/CFT that deformations of the CFT correspond to modifications of the AdS boundary conditions.  For designer gravity theories with a field theory dual, the boundary conditions given by a function $W$ are related to the addition of a multi-trace potential term $\int d^d x~W(\op)$ to the CFT action \cite{witten, bss,Sever:2002fk}, where $\op$ is the operator dual to the bulk scalar.  The effective potential $\V(\op)$ is simply minus the effective lagrangian of the CFT restricted to constant values of $\op$ (and all other sources and currents turned off).  (See, for example, \cite{Papadimitriou:2007sj,Vecchi:2010dd} for discussion of multi-trace deformations and stability from the dual field theory perspective).

Multi-trace deformations of the Dirichlet theory are always irrelevant (the double-trace deformation is classically marginal at the BF bound). However, we will find that in the limit of classical bulk gravity, the theory is in fact UV safe in the following sense:  adding higher power irrelevant deformations of any sign does not destabilize the potential, as it may with irrelevant deformations of the Neumann theory. For double-trace deformations defined at some bare scale, which in particular affect the two-point function at leading order, we find explicitly that the deformed Dirichlet theory flows to a new fixed point in the UV corresponding to the Neumann theory.

This paper is organized as follows.  In section \ref{sec:designergravityreview}, we give a more detailed introduction to designer gravity and review previous work on minimum energy theorems in the Neumann theory.  In section  \ref{dirichletstability},  we derive an energy bound for deformations of the Dirichlet theory and discuss how this relates to properties of the appropriate effective potential.  In section \ref{sec:gluing}, we construct time-symmetric initial data that satisfies our boundary conditions at any value of $\langle \op \rangle$ and whose energy is simply $\V(\langle \op \rangle)$ plus arbitrarily small corrections.  This demonstrates that the object $\V$ is in fact a suitable effective potential. We close with a discussion of our results in section \ref{sec:discussion}.

\section{Designer Gravity Review}
\label{sec:designergravityreview}

In this section, we briefly review the important features of designer gravity theories.  We focus in particular on proofs of the nonperturbative stability of these theories using spinor charge methods.

We consider asymptotically AdS$_{d+1}$ gravity ($d \geq 3$) coupled to a tachyonic scalar field with action
\begin{equation}
\label{theory}
S = \frac{1}{2} \int \, d^{d+1} x \sqrt{ - g} \, [R -
(\nabla \phi)^2 - 2V(\phi)]  \, ,
\end{equation}
where we have set $8\pi G = 1$.  Near $\phi = 0$, we assume that the scalar potential $V(\phi)$
takes the form
\begin{equation}
\label{potential}
V(\phi) = -\frac{d(d-1)}{2\ell_{AdS}^2}+\frac{1}{2} m^{2} \phi^{2} + \ldots \,,
\end{equation}
where $\ell_{AdS}$ is the AdS radius.   It will be convenient to work in units where
$\ell_{AdS} =1$.
 In designer gravity theories,
we restrict to scalar masses $m^2 < 0$ in the range
\begin{equation}
\label{range}
m^2_{BF} \le m^2 < m^2_{BF} +1,
\end{equation}
where the Breitenlohner-Freedman bound for perturbative stability \cite{BF} is
\begin{equation}
m^2 \geq m^2_{BF} =-\frac{d^{\,2}}{4} \,.
\end{equation}

We are interested in metrics which asymptotically approach \cite{Hollands:2005wt,Hertog2005,Amsel:2006uf} the metric of exact AdS
spacetime in global coordinates\footnote{If one is interested in asymptotically planar AdS, one simply needs to take a scaling limit \cite{Faulkner:2010fh} where all scales of the solution are $\gg 1$ in units of the AdS radius. In our case, this is a large $(\alpha,\beta)$ limit.},
\begin{equation}
\label{pureads}
ds^{2} = - \left(1 + r^2 \right)\, dt^2 +
\frac{dr^2}{1 + r^2} + r^2  d \Omega^2_{d-1}\, .
\end{equation}
Here $d \Omega^2_{d-1}$ is the metric on the unit sphere $S^{d-1}$.
For most masses in the range \eqref{range},
the scalar field behaves near the AdS boundary  ($r\to\infty$) as
\begin{equation}
\label{phi}
\phi = \frac{\alpha}{r^{\lambda_{-}}} + \frac{\beta}{r^{\lambda_{+}}}  + \dots \,,
\end{equation}
where
\begin{equation}
\label{roots}
\lambda_{\pm} = \frac{d \pm \sqrt{d^{\,2}+ 4 m^2}}{2} \,,
\end{equation}
and the coefficients $\alpha, \beta$ do not depend on the radial coordinate $r$.
For $m^2 = m^{2}_{BF}$, the roots \eqref{roots} are degenerate and the solution has the asymptotic behavior\footnote{See \cite{Henneaux:2006hk,Amsel:2006uf,resonant} for discussion of additional cases where logarithmic branches may arise.  In general, this can occur when $\lambda_+/\lambda_- = n$, where $n$ is an integer. We will not further consider such cases in this paper (except for $n = 1$).}
\begin{equation}
\label{phibf}
\phi = \frac{\alpha \,\log r}{r^{d/2}} + \frac{\beta}{r^{d/2}} + \dots\,.
\end{equation}
Note that in global AdS we use the radius of the boundary $S^{d-1}$ to define the scale of the logarithm. This means that one should interpret $\log r = \log(r/R_{S^{d-1}}),$ and $R_{S^{d-1}}=\ell_{AdS}=1$ in our units.

In the mass range \eqref{range}, both the $\alpha,\beta$ modes are normalizeable, so in order to have well-defined evolution within a phase space we must impose a
boundary condition at the AdS boundary.  For example, the standard Dirichlet boundary condition is to fix $\alpha = 0$.  Alternatively, one could choose the Neumann boundary condition $\beta = 0$.  More generally, it is sufficient to fix a functional relation between $\alpha$ and $\beta$, which we express as
\begin{equation}
\label{dW}
\beta \equiv \frac{dW}{d\alpha} \, ,
\end{equation}
for some arbitrary smooth function $W(\alpha)$.
Note that a general boundary condition $W$ will break the asymptotic AdS symmetry, but conformal invariance is preserved by the choice
\begin{eqnarray}
\label{cibc}
W(\alpha) &=& k |\alpha|^{d/\lambda_-} \,, \qquad \qquad \qquad m^2 \neq m^2_{BF} \\
\label{bfbc}
W(\alpha) &=& k \alpha^2-\frac{1}{d} \alpha^2 \log |\alpha|\,, \qquad m^2 =m^2_{BF}\label{BFWa}
\end{eqnarray}
for some arbitrary constant $k$.\footnote{It is worth noting that for $m^2 \neq m^2_{BF}$ the Neumann theory $W(\alpha)=0$ preserves the conformal symmetry.   However, this is not true for $m^2 = m^2_{BF}$, since the Neumann boundary condition does not include the logarithmic term in (\ref{BFWa}).  The Dirichlet boundary condition $\alpha=0$ always preserves the conformal symmetry. }

Solitons are nonsingular, horizonless, static, spherically symmetric solutions of the bulk gravity theory.
We expect the minimum energy ground state of a designer gravity theory to be given
by one of these solitons \cite{HH2004, Hertog2005}.  For every choice of $\phi$ at the
 origin, the solutions to the equations of motion behave as in \eqref{phi}
 or \eqref{phibf} for some (constant) values of $\alpha, \beta$.  By scanning different values for $\phi(0)$, we map out a curve in the $\alpha, \beta$ plane\footnote{For certain scalar potentials, this curve may not be
 single-valued, so it does not define a function $\beta_0(\alpha)$.  For example, the known supergravity truncations containing scalars at the BF bound (see e.g., \cite{HM2004,Craps:2007ch}) appear to exhibit this behavior.  We will generally not consider such cases in this work, though we do make some further comments in section \ref{sec:discussion}.}, which we call $\beta_0(\alpha)$.  The solitons consistent with our boundary conditions are then given by the intersection points, $\beta_0(\alpha) = W'(\alpha)$.  Let us now define
\begin{equation}
\label{W0def}
W_0(\alpha) = -\int^\alpha_0 \beta_0(\tilde \alpha) d\tilde \alpha \,,
\end{equation}
and
\begin{equation}
\label{effV}
\V(\alpha) = (\lambda_+-\lambda_-)(W(\alpha) +W_0(\alpha)) \,.
\end{equation}
In general, the function $W_0$ can only be determined numerically, but for small $\alpha$ it can be shown analytically  \cite{Faulkner:2010fh} that
\begin{equation}
W_0 =\frac{\pi}{\lambda_+-\lambda_-}\, \frac{\Gamma^2[\lambda_+/2]\csc[\pi (\lambda_+-\lambda_-)/2]}{\Gamma^2[\lambda_-/2]\Gamma^2[(\lambda_+-\lambda_-)/2]}\,\alpha^2+\ldots
\end{equation}
It was shown in \cite{HH2004} that extrema of $\V$ (denoted $\alpha =
\alpha_*$) correspond to solitons satisfying our boundary conditions,
and further that the value of $\V(\alpha_*)$ gives the total energy  of the
soliton (up to overall volume normalization).

The above statements translate simply to the field theory side. The bulk scalar is dual to an operator $\op$ of conformal dimension $\Delta = \lambda_-$ (which becomes $\Delta = d/2$ when $m^2 = m^2_{BF}$).  Our boundary conditions \eqref{dW} correspond to a deformation of the Neumann theory by adding a term to the action
\begin{equation}
\Delta S_{CFT} = - \int d^d x ~ W(\op).
\end{equation}
Note that for masses in the range \eqref{range}, a double-trace deformation $\op^2$ is a relevant operator (or marginal when $m^2 = m^2_{BF}$). The function $\V$ is simply the effective potential for the operator, which is minus the effective action restricted to constant field configurations (see e.g., \cite{Papadimitriou:2007sj}),
\begin{equation}
\V(\alpha) = - \Gamma[\op(x^\mu)=\alpha] \,.
\end{equation}
Every soliton corresponds to an extremum of $\V$ with $\langle \op
\rangle = \alpha_*$, and the energy of the state is simply $\int d^dx ~ \V(\alpha_*)$.
Based on this interpretation, it was conjectured in \cite{HH2004} that the theory would be stable if $\V$ admits a global minimum.  We now briefly review previous work on proving this conjecture.

\subsection{Stability in the Neumann Theory}
\label{NeumannStability}

The minimum energy bound is derived
following a Witten-Nester style proof \cite{Witten:1981mf,Nester:1982tr,Gibbons:1983aq,Boucher:1984yx,Townsend:1984iu}, which makes use of the spinor charge
\begin{equation}
\label{charge}
Q = \int_{C} \ast {\bf B} \,, \quad B_{cd} = \bar{\psi} \gamma_{[c}\gamma_d\gamma_{e]} \widehat{\nabla}^e \psi + \textrm{h.c.}\, ,
\end{equation}
where $C =\partial \Sigma$ is a surface at spatial infinity that bounds a
spacelike surface $\Sigma$.  The covariant derivative is
\begin{equation}
\widehat{\nabla}_a \psi = \nabla_a \psi +
\frac{P(\phi)}{\sqrt{2(d-1)}} \gamma_a \psi \,,
\end{equation}
where the ``Witten spinor'' $\psi$ is required to satisfy a spatial Dirac equation $\gamma^i \widehat{\nabla}_i \psi = 0$ and to asymptotically approach a Killing spinor of exact AdS (see e.g., \cite{Amsel:2007im}). Using standard manipulations, it can be shown that $Q \geq 0$ if the
``superpotential'' $P$ satisfies
\begin{equation}
 \label{vtop}
 V(\phi) = (d-1) \left(\frac{dP}{d\phi}\right)^2 -d P^2 \, .
\end{equation}
The energy bound is then obtained by relating the spinor charge to the physical energy $E$.  This depends crucially on the global existence of a (real) one-parameter family of superpotentials $P_s$ \cite{Amsel:2007im,Faulkner:2010fh}.  Generally, a solution to \eqref{vtop} exists up to some critical value $s_c >0$, and we refer to this as the critical superpotential, $P_c$.

We first take $m^2 \neq m^{2}_{BF}$.   Assuming that the appropriate superpotential exists, evaluation of the spinor charge yields \cite{Amsel:2006uf,Amsel:2007im,Faulkner:2010fh}
\begin{equation}
\label{sbound}
E \geq (\lambda_+ - \lambda_-)\oint \left[W+\frac{\lambda_-}{d} \, s_c |\alpha|^{d/\lambda_-}\right] \,,
\end{equation}
where the integral is over the unit sphere $S^{d-1}$.  The explicit form of the total conserved energy $E$ may be found in, e.g.~\cite{Amsel:2006uf}.

Using scaling arguments, one can show \cite{Faulkner:2010fh} that for large $\alpha$, the spherical soliton approaches a planar, boost-invariant solution (sometimes called a ``fake supergravity'' domain wall \cite{DeWolfe:1999cp, Freedman:2003ax}), and in this asymptotic regime
\begin{equation}
\label{neumannsi}
W_0(\alpha) = \frac{\lambda_-}{d} \,s_c |\alpha|^{d/\lambda_-} +\ldots\,
\end{equation}
The fake supergravity solution (which is an asymptotically \emph{planar} AdS solution) in fact saturates the energy bound \eqref{sbound}.
It follows that if $\V(\alpha)$ is bounded, the right hand side of \eqref{sbound} is bounded. This proved the stability conjecture in \cite{HH2004}.  A similar result has been proven when $m^2 = m^{2}_{BF}$ in \cite{Amsel:2011km} (see Appendix~\ref{appendix}).

\begin{figure}[h!]
  \centering
  \includegraphics[scale=.76]{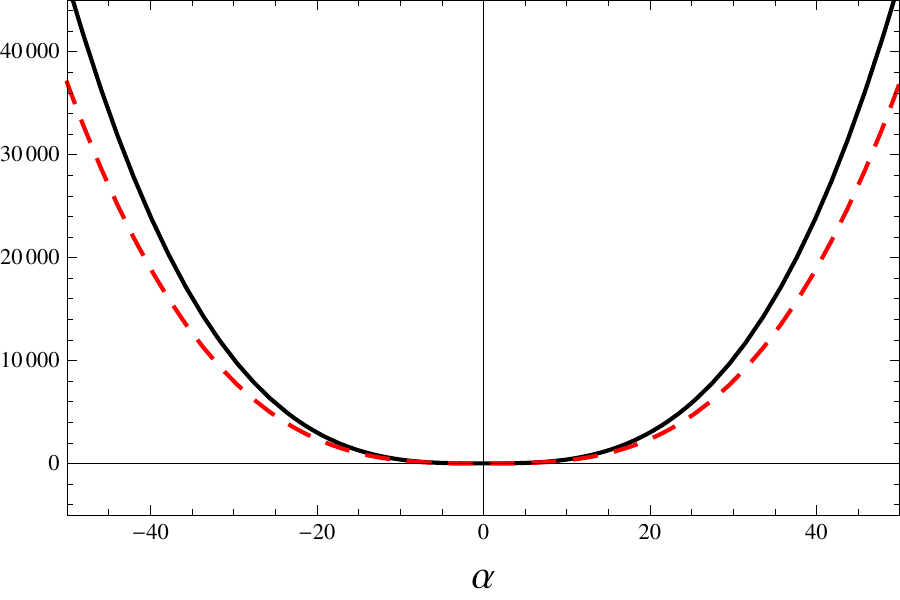}
  \hspace{.5cm}
  \includegraphics[scale=.76]{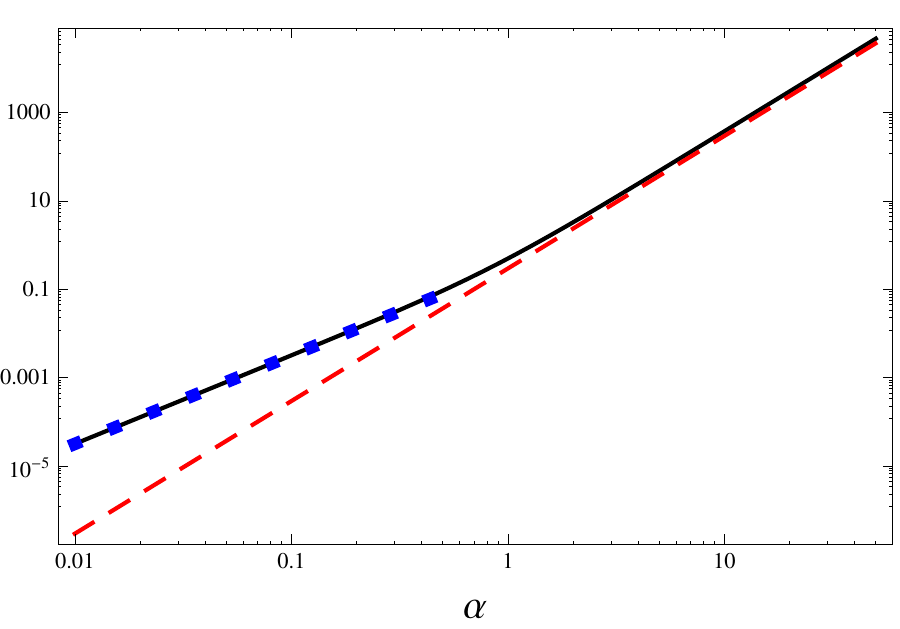}
  \caption{
 The Neumann effective potential function $W_0(\alpha)$ in global AdS (solid black) and Poincar\'e AdS (dashed red). These plots are for the case of $d=3,~\lambda_-=1,V=-3-\phi^2+\phi^4/2.$ In the log plot (right), we also include the small $\alpha$ analytic answer, $W_0 = \frac{\alpha^2}{\pi}+\ldots$ (dotted blue). Note that here it is sufficient to simply plot $W_0$, as the contribution of $W(\alpha)$ to $\V$ is trivially additive. This is not the case for the Dirichlet theory.
  \label{fig:neu_potential}}
\end{figure}


\section{Stability in the Dirichlet Theory}
\label{dirichletstability}

In this section, we turn to studying deformations of the Dirichlet theory $\alpha = 0$.  We derive the appropriate energy bound for these cases and analyze the conditions under which the theory is stable.

The previous section considered boundary conditions of the form $\beta = \beta(\alpha)$, corresponding to a deformation of the Neumann theory $\beta = 0$.  An equally valid boundary condition that also leads to a single consistent phase space is given by enforcing $\alpha = \alpha(\beta)$.   This corresponds to a deformation of the Dirichlet theory.  Once again, it is useful to parametrize this in terms of some arbitrary function $W(\beta)$ as
\begin{equation}
\label{dirichletbc}
\alpha = \frac{dW}{d\beta} \,.
\end{equation}
This boundary condition corresponds to a deformation of the field theory by a term $W(\op)$, where the operator $\op$ now has conformal dimension $\Delta = \lambda_+$.  Note that for masses in the range \eqref{range}, a deformation $\op^n$ for $n \geq 2$ is an irrelevant operator (or marginal if $n = 2$ and $m^2 = m^2_{BF}$).  For now we consider general masses $m^2_{BF} < m^2 < m^2_{BF}+1$.  The case where the BF bound is saturated will be treated separately in Appendix~\ref{appendix}.

\subsection{The Energy Bound}

The energy bound for the Dirichlet theory can be derived following the same procedure given in section~\ref{NeumannStability};  the only difference is the scalar boundary condition \eqref{dirichletbc}.
The final result is
\begin{equation}
\label{bound0}
E \geq (\lambda_+-\lambda_-)\oint \left( \alpha \beta - W(\beta) +\frac{s_c \lambda_-}{d} \, |\alpha|^{d/\lambda_-} \right) \,,
\end{equation}
where we emphasize that in this expression $\alpha$ is regarded as a function of $\beta$ due to the boundary condition.  This means that every term in \eqref{bound0} depends explicitly on the function $W$, in contrast to the general Neumann case\footnote{When $s_c = 0$ (which occurs for example in the supergravity truncation studied in \cite{HH2004}), we see from \eqref{sbound} that stability of the Neumann theory depends simply on whether the boundary condition $W(\alpha)$ has a global minimum.  For the Dirichlet case with $s_c = 0$, stability depends on the more complicated expression $\beta W'(\beta)-W(\beta)$, though at least for the power-law boundary conditions \eqref{simplebc} this is again just proportional to $W(\beta)$.}.

The deformed Dirichlet theory will be stable if the quantity on the right hand side of \eqref{bound0}
is bounded from below.  To proceed with the analysis, we assume a simple form for the deformation,
\begin{equation}
\label{simplebc}
W(\beta) = f\, \beta^n \quad \implies \alpha(\beta) = n f \beta^{n-1} \,,
\end{equation}
for $n$ a non-negative integer and $f$ a constant.  In theories with a CFT dual, such polynomial deformations $W\sim\op^n$ are typically of most interest.

The energy bound then becomes
\begin{equation}
\label{Ebound}
E \geq (\lambda_+-\lambda_-)\oint\left((n-1)f \beta^n +\frac{s_c \lambda_- n|f|^{d/\lambda_-}}{d} \, |\beta|^{(n-1)d/\lambda_-} \right) \,.
\end{equation}
For $n = 0$ or $1$, the inequality \eqref{Ebound} takes the form $E \geq constant$, so the theory is stable.  For $n \geq 2$, the second term on the right hand side of \eqref{Ebound} dominates at large $\beta$.  Since the coefficient of this term is positive, there is always a global minimum.
We conclude that for any $n,f$, the deformed Dirichlet theory is stable.

\subsection{The Effective Potential}

Following the arguments in the Neumann case, we may attempt to define an analogous effective potential along the lines of (\ref{W0def},\ref{effV}),
\begin{equation}
\label{Vtilde}
\tV(\beta)=(\lambda_+-\lambda_-)\left(W_0(\beta)-W(\beta)\right) \,,
\end{equation}
where
\begin{equation}
W_0(\beta)=+\int_0^\beta \alpha_0( \tilde \beta)d\tilde\beta
\end{equation}
and $\alpha_0(\beta)$ is the soliton curve (Note that this is a different function than \eqref{W0def}, which takes $\alpha$ as the argument; to avoid confusion we shall always write the argument ($\alpha$ or $\beta$) explicitly).
Once again, this satisfies the properties that extrema of $\tV$ give the solitons satisfying our boundary conditions and the value of $\tV$ at the extremum is the energy of the corresponding soliton.

For large $\beta$, the solitons approach the scale-invariant form
\begin{equation}
W_0(\beta) =- \frac{\lambda_+}{d s_c^{\lambda_-/\lambda_+}} |\beta|^{d/\lambda_+} \,+\ldots,
\end{equation}
which may be obtained from \eqref{W0def} and \eqref{neumannsi}.  Hence, the naive effective potential approaches
\begin{equation}
\tV =(\lambda_+-\lambda_-)\left( -f \beta^n- \frac{\lambda_+}{d s_c^{\lambda_-/\lambda_+}} \,|\beta|^{d/\lambda_+}+\ldots \right)\,.
\end{equation}
Clearly this expression does not match the right hand side of the energy bound \eqref{Ebound}, which is a major difference from the Neumann case.  In particular, there are cases for which $\tV$ has no global minimum (for example, $W(\beta) =  f \beta^2,~f> 0$) but, as stated above, $E$ is always bounded from below.  This suggests that $\tV$ is not the correct effective potential for the Dirichlet theory (as indicated by the tilde notation).

Let us instead consider
\begin{equation}
\label{VeffD}
 \V(\beta) = (\lambda_+-\lambda_-) \left[ \alpha(\beta) \beta - W(\beta) +W_0(\alpha(\beta)) \right] \,,
\end{equation}
which is simply the ``Legendre transform''  $W(\alpha) \to \alpha \beta - W(\beta)$ of the Neumann effective potential \eqref{effV}. First, note that when a soliton satisfies our boundary conditions (at some $\beta = \beta_*$), we have
\be
\V'(\beta_*) = (\lambda_+-\lambda_-) \left[ \alpha(\beta_*) - W'(\beta_*) + \alpha'(\beta_*) \left\{W_0'(\alpha(\beta_*)) + \beta_* \right\} \right]=0,
\ee
and at these extrema $\V(\beta_*)$ gives the energy of the soliton. Furthermore, using the asymptotic scaling behavior (\ref{neumannsi}) we have in the planar limit,
\begin{equation}
 \V(\beta) = (\lambda_+-\lambda_-) \left( \alpha(\beta) \beta - W(\beta) +\frac{s_c \lambda_-}{d} \, |\alpha(\beta)|^{d/\lambda_-} +\ldots\right) \,,\label{scaling_dir_bound}
\end{equation}
which is just the expression that appears in the energy bound \eqref{bound0}, as was the case for the Neumann theory. It follows that once again, the planar fake supergravity solutions saturate the energy bound.  We have therefore found a good candidate effective potential, which gives the mass of static solutions at its extrema and in the planar (large $\beta$) limit coincides with the energy bound.  We emphasize that this was not the case for $\tV$. This asymptotic agreement of $\V(\beta)$ and the RHS of \eqref{Ebound} tells us stability for the Dirichlet theory is guaranteed by boundedness of $\V(\beta)$. Following the original designer gravity conjecture, we will also conjecture that the minimum energy solution is the spherical soliton associated with the minimum of \eqref{VeffD}. Further evidence that we should interpret \eqref{VeffD} as the effective potential for the deformed Dirichlet theory will be provided in the next section.

\begin{figure}[h!]
  \centering
  \includegraphics[scale=.76]{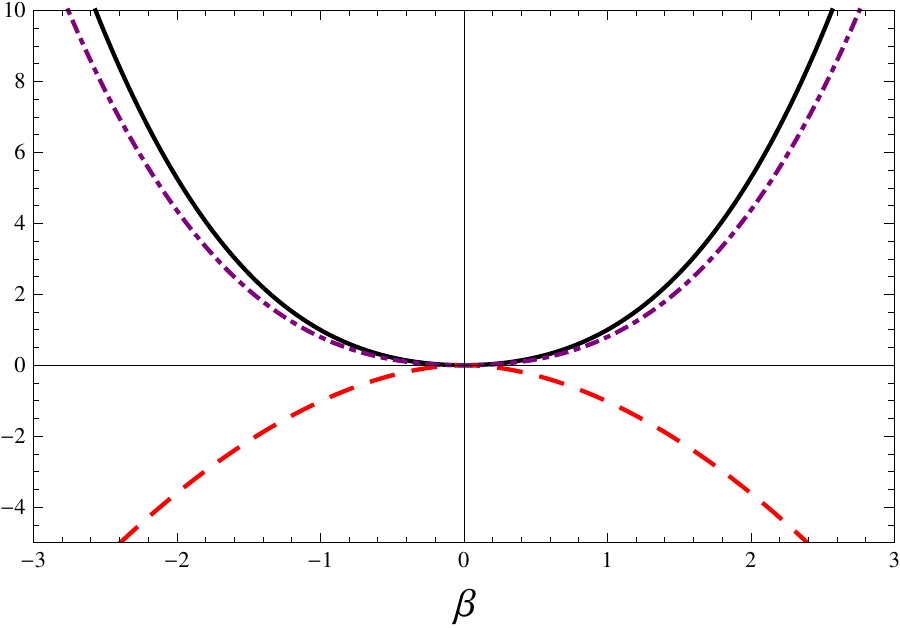}
  \hspace{.5cm}
  \includegraphics[scale=.76]{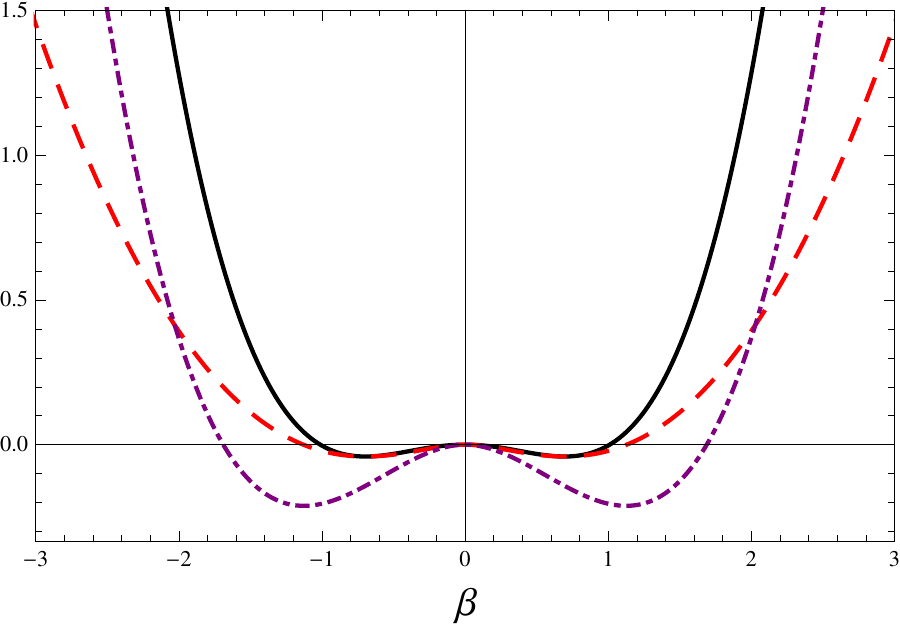}
  \caption{
 The Dirichlet effective potential function $\V$ (solid black), the incorrect potential function $\tV$ (dashed red), and the quantity on the right hand side of the energy bound \eqref{bound0} (dash-dot purple).  Here we have chosen linear boundary conditions $\alpha = W'(\beta) = f \beta$, with $f>0$ (left) and $f<0$ (right) in global AdS. These plots are for the case of $d=3,~\lambda_+=2,V=-3-\phi^2+\phi^4/2, f=\pm 1$, which has $s_c=0.89$.  Note that $\tV$ and $\V$ agree at extrema, and that the true effective potential is bounded from below for either sign of $f$.  \label{fig:dir_potential}}
\end{figure}


\section{The Off-Shell Effective Potential}
\label{sec:gluing}
We have argued above that the effective potential $\V$ is a good effective potential in that its value at extrema gives the energy of corresponding static configurations.  In this section, we will improve upon this by showing that both $\V(\alpha)$ and $\V(\beta)$ also give the energy of nearby non-static configurations.

 In mean field theory, we can consider evaluating the effective action on a constant field state and construct the usual effective potential,
\be
\V(\varphi_0) = - (\mathrm{Volume})^{-1}\Gamma[\varphi(x) = \varphi_0] \,.
\ee
This is a measure of the energy of a mean field state even when $\V' \neq 0$, so that it is not a static state. We wish to construct a similar object holographically.  To do so, we will construct  initial data which asymptotically satisfies our Dirichlet (Neumann) boundary conditions $\alpha = W'(\beta)$  ($\beta = W'(\alpha)$) for arbitrary $\beta$ ($\alpha$).  This data is matched to an interior solution that is a constant time slice of an AdS soliton or domain wall, where the approximate asymptotic interior $\alpha_*, \beta_*$ coefficients are related via the soliton curve $\beta_0(\alpha)$. Since the slice of the soliton/domain wall is time-symmetric, it is natural to consider the case where the exterior data is time-symmetric as well.

\subsection{Energy of Initial Data}
Let us review the $(d+1)$ Hamiltonian decomposition of $AdS_{d+1}$. We express the bulk metric as
\be
{}^{(d+1)}g_{ab} = {}^{(d)}h_{ab} - n_a n_b \,,
\ee
where $n^a$ is the unit normal to the spacelike hypersurface.  We consider time-symmetric initial data with zero shift,
\be
N^i=0,~n^a\nabla_a \phi=0,~K_{ij}=0.
\ee
There is one constraint equation, which can be written as
\be
\frac{1}{2} {}^{(d)}R = \frac{1}{2}h^{ij} \partial_i \phi \partial_j \phi+V(\phi).\label{eq:constraints_genl}
\ee
If we assume a rotational symmetry in the non-radial directions of the spatial metric, we can use the simple ansatz
\be
\phi = \phi(r),~ds^2(h) = r^2 d\Omega_k^2 + \frac{dr^2}{f(r)},
\ee
where $d\Omega_k^2$ is the unit metric on a $S^{d-1}~(k=1)$,~$\mathbb{R}^{d-1}~ ( k=0)$,~$\mathbb{H}^{d-1}~(k=-1).$ The constraint equation then becomes
\be
\frac{k}{2}\frac{(d-2)(d-1)}{r^2} - \frac{d-1}{2} \frac{(r^{d-2} f)'}{r^{d-1}} = V(\phi) + \frac{1}{2} f \phi'^2,\label{eq:constraints_sym}
\ee
where primes denote derivatives with respect to $r$. If our initial profile for $\phi$ has the usual asymptotics
\be
\phi = \frac{\alpha}{r^{\lambda_-}} + \frac{\beta}{r^{\lambda_+}}+\ldots,\label{eq:phi_asympt}
\ee
then solving (\ref{eq:constraints_sym}) for large $r$ yields
\be
f = r^2 + k +\frac{\lambda_-}{d-1} \frac{\alpha^2}{r^{2-2\lambda_-}} - \frac{M_0}{r^{d-2}} + \ldots \,,\label{eq:metric_asympt}
\ee
where $M_0$ is an undetermined constant.  For this time- and rotationally-symmetric initial data, the differential Hamiltonian is
\be
\delta H = - \oint_{\Omega_k } N f^{-1/2} \left( \frac{d-1}{2}r^{d-2} ~\delta f + r^{d-1} f ~\phi' ~ \delta \phi \right),\label{eq:sym_diff_hamiltonian}
\ee
where for our asymptotically $AdS_{d+1}$ solutions we will normalize the lapse as $N=r+\ldots$, following \eqref{pureads}.  Substituting (\ref{eq:phi_asympt},~\ref{eq:metric_asympt}) into (\ref{eq:sym_diff_hamiltonian}) we find
\be
\frac{E_\alpha}{V_k} = \frac{d-1}{2}M_0 + \lambda_- \alpha \beta + (\lambda_+ -\lambda_-)W(\alpha)\,,~ \quad(\mathrm{Neumann}),\label{eq:neumann_en}
\ee
\be
\frac{E_\beta}{V_k} = \frac{d-1}{2} M_0 + \lambda_+ \alpha \beta - (\lambda_+ -\lambda_-)W(\beta)\,, ~\quad(\mathrm{Dirichlet}).\label{eq:dirichlet_en}
\ee

\subsection{Gluing Solitons}

The soliton solution is of the form (\ref{eq:phi_asympt},\ref{eq:metric_asympt}) with $k=1$ and $\beta = \beta_0(\alpha)$.  Comparing \eqref{eq:neumann_en} to \eqref{effV}  (or \eqref{eq:dirichlet_en} to \eqref{Vtilde}) we see that the bare mass coefficient in $h^{rr}$ is related to $W_0(
\alpha)$ as
\be
\frac{d-1}{2}M_0= (\lambda_+-\lambda_-)W_0(\alpha)- \lambda_-\alpha \beta .
\ee

For simplicity, let us first consider the case $d = 3, \lambda_- = 1, V(\phi) = -3-\phi^2+\lambda \phi^4/4!$.  Then the interior solution is
\begin{eqnarray}
\phi_s &=&\frac{\alpha_*}{r} + \frac{\beta_*}{r^2}+\frac{(3+\lambda)\alpha_*^3}{12r^3}+\ldots\\
f_s(r) &=& r^2+1 + \frac{\alpha_*^2}{2}-\frac{M_{s}}{r}+\frac{4 \alpha_*^2+8\beta_*^2+(4+\lambda)\alpha_*^4}{8 r^2}+\ldots  \,.
\end{eqnarray}
and the exterior solution is
\begin{eqnarray}
\phi_{ext} &=&\frac{\alpha}{r} + \frac{\beta}{r^2}\\
f_{ext}(r) &=& r^2+1 + \frac{\alpha^2}{2}-\frac{M_{ext}}{r}+\frac{12 \alpha^2+24\beta^2+(6+\lambda)\alpha^4}{24 r^2}+\ldots \,.
\end{eqnarray}
At some $r = r_* \gg |\alpha_*|$, we impose the matching conditions
\be
f_{ext} (r_*) = f_s(r_*),~\phi_{ext} (r_*) = \phi_s(r_*)
\ee
which gives
\begin{equation}
M_{ext} = W_0(\alpha)-\alpha \beta +\frac{(\beta_0(\alpha)-\beta)^2}{2 r_*}+\ldots \,.
\end{equation}

For general dimension and scalar mass, this becomes
\begin{equation}
M_{ext} = \frac{2}{d-1}\left((\lambda_+-\lambda_-)W_0(\alpha)-\lambda_-\alpha \beta\right) +\frac{(\lambda_+-\lambda_-)(\beta_0(\alpha)-\beta)^2}{(d-1) r_*^{\lambda_+-\lambda_-}}+\ldots \,.
\end{equation}
Substituting this into the energy formula \eqref{eq:neumann_en}, we find for deformed Neumann theory that
\begin{equation}
\frac{E_\alpha}{V} = (\lambda_+-\lambda_-) \left(W(\alpha)+ W_0(\alpha) +\frac{(\beta_0(\alpha)-\beta)^2}{2r_*^{\lambda_+-\lambda_-}}+\ldots\right)\label{neu_glue_energy}
\end{equation}
Similarly, for the deformed Dirichlet theory, \eqref{eq:dirichlet_en} gives
\begin{equation}
\frac{E_\beta}{V} = (\lambda_+-\lambda_-)\left(\alpha\beta - W(\beta)+
 W_0(\alpha(\beta))
 +\frac{(\beta_0(\alpha(\beta))-\beta)^2}{2r_*^{\lambda_+-\lambda_-}}+\ldots\right) \,.\label{dir_glue_energy}
 \end{equation}
We have therefore found that our gluing construction gives solutions which satisfy our boundary conditions for any desired value of $\alpha$ or $\beta$, with energy arbitrarily close to the effective potential $\V(\alpha)$ or $\V(\beta)$. Further, the  subleading corrections to both (\ref{neu_glue_energy}, \ref{dir_glue_energy}) are positive definite.  This gives strong evidence for the second conjecture of \cite{HH2004}, namely that in global AdS the minimum energy solution is the spherical soliton associated with the global minimum of $\V$.
\subsection{Gluing Planar Solutions}

The gluing story for asymptotically planar AdS systems is very similar to what we just studied above. However, in the planar limit, one no longer has a nice smooth polar coordinate origin at $r=0$, so one must be careful about what interior solution we choose. However, if we simply take the $\alpha \gg 1$ limit, it has been shown that solitons approach scale-invariant planar domain walls which saturate the energy bounds (\ref{sbound},~\ref{bound0}) \cite{Faulkner:2010fh}. The domain wall solution is given in terms of the critical superpotential $P_c$,
\be
k=0,~N=r,~ f = \frac{2 r^2 P(\phi(r))^2}{d-1},~\phi'(r) = -\frac{(d-1) P_{,\phi}(\phi(r))}{rP(\phi(r))}.
\ee
The one constant of integration fixes the value of $\alpha$, and the scaling symmetry of the solution guarantees that
\be
\beta = - s_c  \alpha |\alpha|^{(\lambda_+-\lambda_-)/\lambda_-} ,~M_0 = \frac{\lambda_+\lambda_-}{d(d-1)}4 s_c |\alpha|^{d/\lambda_-}.\ee
These solutions may have null singularities at $r=0$ (or simply a Poincar\'e horizon) depending on the details of the potential and therefore $P_c$.  Nevertheless, it was argued in \cite{Faulkner:2010fh} that they are valid solutions to consider because they are limiting cases for both global solitons and finite temperature planar black holes with scalar hair.

Let us once again consider the simple case $d = 3, \lambda_- = 1, V(\phi) = -3-\phi^2+\lambda \phi^4/4!$.  Then there is a one-parameter family of solutions labeled by $\alpha_*$,
\be
f_s(r) = r^2 + \frac{\alpha_*^2}{2}-\frac{4 s_c |\alpha_*|^{3}}{3r}+\frac{(4+8s_c^2+\lambda)\alpha_*^4}{8 r^2}+\ldots\nonumber
\ee
\be
\phi_s =\frac{\alpha_*}{r} - \frac{s_c \alpha_* |\alpha_*|}{r^2}+\frac{(3+\lambda)|\alpha_*|^3}{12r^3}+\ldots \,.
\ee
We will simply take this solution, cut it off at some $r_* \gg |\alpha_*|$ and match it to a generic solution of the form (\ref{eq:phi_asympt},~\ref{eq:metric_asympt}),
\be
f_{ext.} (r_*) = f_s(r_*),~\phi_{ext.} (r_*) = \phi_s(r_*).
\ee
Solving for $M_0$ perturbatively in $1/r_*$, we find
\be
M_0 = \frac{s_c |\alpha|^{3}}{3}-\alpha \beta +\frac{(s_c\alpha|\alpha|+\beta)^2}{2 r_*}+\ldots \,.
\ee

For general dimension and scalar mass, the corresponding result is
\be
M_0 = \frac{2\lambda_-}{d-1} \left(\frac{(\lambda_+-\lambda_-) s_c |\alpha|^{d/\lambda_-}}{d}-\alpha \beta \right)+\frac{\lambda_+-\lambda_-}{d-1}\frac{(s_c\alpha|\alpha|^{(\lambda_+-\lambda_-)/\lambda_-}+\beta)^2}{r_*^{\lambda_+-\lambda_-}}+\ldots
\ee
Substituting this into the energy formula \eqref{eq:neumann_en}, we find for deformed Neumann theory that
\be
\frac{E_\alpha}{V} = (\lambda_+-\lambda_-) \left(W(\alpha)+ \frac{\lambda_- s_c|\alpha|^{d/\lambda_-}}{d} +\frac{(s_c\alpha|\alpha|^{(\lambda_+-\lambda_-)/\lambda_-}+\beta)^2}{2r_*^{\lambda_+-\lambda_-}}+\ldots\right),
\ee
which can come arbitrarily close to the bound (\ref{sbound}). Similarly, for the deformed Dirichlet theory, \eqref{eq:dirichlet_en} gives
\be
\frac{E_\beta}{V} = (\lambda_+-\lambda_-)\left(\alpha\beta - W(\beta)+
 \frac{\lambda_-s_c|\alpha|^{d/\lambda_-}}{d}
 +\frac{(s_c\alpha|\alpha|^{(\lambda_+-\lambda_-)/\lambda_-}+\beta)^2}{2r_*^{\lambda_+-\lambda_-}}+\ldots\right),
 \ee
which can come arbitrarily close to (\ref{bound0}).  Note that, as before, the leading correction to the energy of the planar domain wall is manifestly positive, which is consistent with the spinor charge energy bounds for planar AdS solutions given in the previous sections. We could have also arrived at these energy formulae by simply taking the planar scaling limit \eqref{neumannsi} of our soliton result.


\section{Discussion}\label{sec:discussion}

In this paper, we have proved a minimum energy theorem \eqref{bound0} for asymptotically AdS spacetimes containing scalar fields with general boundary conditions.  Whereas all previous proofs restricted to boundary conditions that are deformations of the Neumann theory $\beta = \beta(\alpha)$, we have given a complete treatment of stability for  boundary conditions that are deformations of the Dirichlet theory $\alpha = \alpha(\beta)$.  Similarly to the Neumann case, the result can be stated in terms of an effective potential:  If the function \eqref{VeffD} has a global minimum, then the energy is bounded from below.
In particular, we showed that holographic theories can be stable under deformations which are power-counting irrelevant, contrary to what one might have expected.  We now conclude by discussing several aspects of these results.

\subsection{When $s_c$ Is Not Positive}

As mentioned above, there are examples of scalar potentials corresponding to $s_c=0$, some of which arise in known supergravity truncations.
This often occurs when the soliton curve $\beta_0(\alpha)$ is not invertible (see e.g.~figure 2 of \cite{Faulkner:2010fh}). In this case, the stability analysis is slightly modified. For Neumann-deformed theories, this was discussed in \cite{Faulkner:2010fh}.  The results for the Dirichlet case are similar: without the stabilizing term proportional to $s_c$ in \eqref{scaling_dir_bound}, stability depends purely on $W(\beta)$.

Another possibility is $s_c <0$, which implies that no power-law boundary conditions can stabilize the Dirichlet theory.  However, the Neumann theory can still be stabilized for certain boundary conditions.

In some theories, $\beta_0(\alpha)$ is not even single-valued. For instance, there are cases where the curve has the behavior $\alpha \rightarrow const.$ as $\beta \rightarrow \infty$ (see e.g.~figure 7 of \cite{Amsel:2011km}). This implies $s_c  = -\infty$ (using the language of  \cite{Amsel:2007im}  the critical super potential is $P_+$). This suggests that no boundary conditions other than undeformed Dirichlet ($\alpha = 0$) lead to a stable theory.

\subsection{Double-trace Deformations}
Let us study more closely the case of double-trace deformations, when the boundary conditions are linear. It is well-known that turning on a relevant double-trace deformation in a Neumann quantized theory causes a flow to the Dirichlet theory in the IR.  Because the deformation is relevant, the UV theory is simply the undeformed Neumann theory \cite{witten}. This is most easily seen in terms of correlators (following e.g.~\cite{Faulkner:2010gj,Iqbal:2011aj}).  Labeling the Neumann (Dirichlet) retarded Green's function $G^-(p)~(G^+(p))$, one can show that under a relevant (irrelevant) deformation $W(\op) = f \op^2$ we have
\be
G^-_f(p) = \frac{1}{1/G^-(p)+2f}+O(1/N^2)~\quad \mathrm{(Neumann)},
\ee
\be
G^+_f(p) = \frac{1}{1/G^+(p)+2f}+O(1/N^2)~\quad \mathrm{(Dirichlet)},
\ee
where $p^2 = -(w+i\epsilon)^2+\vec{p}^{\,2}$. In the infrared,
\begin{eqnarray}
G^-_f(p)  &=& \frac{1}{2f}-\frac{1}{4f^2G^-(p) } +\ldots \sim G_+(p),\\
 G^+_f(p) &=& G^+(p)+\ldots,
\end{eqnarray}
so both theories flow to the Dirichlet theory. Furthermore, in the ultraviolet,
\begin{eqnarray}
G^-_f(p)  &=& G^-(p)+\ldots,\\
G^+_f(p) &=& \frac{1}{2f} - \frac{1}{4f^2G^+(p)}+\ldots \sim G^-(p),
\end{eqnarray}
and so we find that under a double-trace deformation, both the Neumann and Dirichlet theories have a nice UV completion as the undeformed Neumann theory.

This is surprisingly similar to what happens in large $N$ vector models \cite{Parisi:1975im}. Consider an $O(N)$ spin-zero vector theory in 2+1 dimensions at leading order in large $N$. In the UV, the operator $\vec \phi \cdot \vec \phi$ has dimension $1$, and flows to dimension $2$ in the IR. Meanwhile, a vector of fermions $\vec \psi \cdot \vec \psi$ has constructive dimension $2$, but due to its fermionic nature, the corresponding beta function has the opposite sign and the operator flows to dimension 1 in the UV. In standard supergravity truncations to $AdS_4$ with $m^2 = -2$, we have scalar bilinears of the form $\mathrm{Tr}\, \phi^2$ with dimension 1, and fermion bilinears of the form $\mathrm{Tr} \,\psi^2$ with dimension 2, so our holographic model reproduces behavior similar to the vector model result.

\subsection{Other Gluing Attempts}
In planar AdS, the energy of our glued solutions saturates the spinor energy bound. This is not the case in global AdS, but the fact that the leading correction to the energy is manifestly positive highly suggests that the minimum energy solution is the spherical soliton associated with the global minimum of $\V$.   However, it is certainly not proven that one cannot construct a different solution (or simply initial data) that is smooth and has an energy in the range $\min \V > E > \min E_Q$, where $E_Q$ is the spinor charge bound (\ref{sbound}) or \eqref{bound0}. It would be interesting to investigate if there exist other gluing constructions with energy below  $\min \V$, which would disprove this part of the designer gravity conjecture. Other known examples of similar gluing constructions (e.g.~\cite{Hubeny:2004cn}) do not even saturate $\V$, as their interior solution is not chosen to minimize the energy.

\subsection{Time Evolution of Initial Data}
Since we have provided initial data and boundary conditions, we may study the time evolution of our glued initial data. Within the causal diamond of the region $r<r_*$, the solution is simply the static soliton, but outside (unless we are at an extremum of $\V$) the solution will simply collapse to a black hole. Depending on the details of the initial value of $\alpha$ (or $\beta$)  as well as the boundary conditions, the system may evolve to a static black hole with or without hair. It would be very interesting to study the evolution of this initial data as a strongly coupled analog to the simple case of a scalar field in a $\V(\phi)$ potential. Indeed, one obvious signature of the strong coupling is the fact that the system always thermalizes.

\subsection{Finite $N$ Effects}
It needs to be emphasized that our result is only valid at infinite $N$. It is of course possible that one-loop corrections will completely destroy the UV stability of our theory. However, there are known cases where irrelevant deformations are less malignant, and for instance only renormalize the AdS radius  \cite{Gubser:2002zh}. It would therefore be of great interest to study the general effects of radiative corrections on these stability theorems.


\section*{Acknowledgments}
It is a pleasure to thank Massimo Porrati for useful discussions.  MMR is supported by the Simons Postdoctoral Fellowship Program.

\appendix
\section{Stability at the BF Bound}
\label{appendix}

In this appendix, we discuss energy bounds when the BF bound is saturated, i.e.~$m^2 = m^2_{BF}$.

For deformations of the Neumann theory with boundary conditions $\beta = W'(\alpha)$, the energy bound is
\begin{equation}
\label{bfboundfinal}
E \geq \oint\left[W(\alpha) +k_c
\alpha^2 +\frac{1}{d} \alpha^2 \log |\alpha| \right] \,,
\end{equation}
where once again we require the global existence of the appropriate superpotential, and $k_c$ is a constant associated with the critical solution.  For large $\alpha$, the soliton curve behaves as
\begin{equation}
\label{bfW0large}
W_0(\alpha) = k_c \alpha^2 + \frac{1}{d} \alpha^2 \log |\alpha| \,,
\end{equation}
which matches the behavior of the planar fake supergravity solution.  It follows that when the effective potential $\V = W+W_0$ has a global minimum, the energy is bounded from below \cite{Amsel:2011km}.

 For deformations of the Dirichlet theory\footnote{As noted above, the Neumann theory $\beta = 0$ is not conformal when the BF bound is saturated.  Since the Dirichlet theory $\alpha = 0$ is conformal, one might consider it more natural (in the context of AdS/CFT) to instead study deformations of this theory.} with boundary conditions $\alpha = W'(\beta)$, the results are qualitatively similar to those away from the BF bound.  In this case, the energy bound becomes  \cite{Amsel:2011km}
\begin{equation}
\label{bound0bf}
E \geq  \oint \left(\alpha \beta - W(\beta) +k_c\alpha^2 +\frac{1}{d} \alpha^2 \log|\alpha| \right)\,,
\end{equation}
so for the boundary conditions \eqref{simplebc}, we have
\begin{equation}
\label{Ebound2}
E \geq \oint\left( (n-1)f \beta^n +\left(\frac{\log|n f|}{d}+k_c \right) n^2 f^2 \beta^{2(n-1)}+\frac{n^2 f^2 (n-1)}{d} \, \beta^{2(n-1)} \log| \beta| \right)\,.
\end{equation}
For $n = 0,1$,  the energy bound reads $E \geq constant$, so the theory is stable.
For $n \geq 2$, the last term on the right hand side of \eqref{Ebound2} dominates at large $\beta$.  Since the coefficient of this term is positive, there is always a global minimum.
We conclude once again that for any $n,f$, the deformed Dirichlet theory is stable.

For $m^2 = m^2_{BF}$, the scale-invariant soliton curve leads to
\begin{equation}
W_0(\beta) = -\frac{d \beta^2}{8 w_p\left(\pm \frac{d \beta}{2 (e^{k_0} \Lambda)^{d/2}}\right)^2} \, \left(1+2 w_p\left(\pm \frac{d \beta}{2 (e^{k_0} \Lambda)^{d/2}}\right) \right) \,,
\end{equation}
where $k_0 = -(2 k_c + 1/d)$ and $w_p(z)$ is the generalized Lambert function \cite{Amsel:2011km}.  Hence, the naive effective potential \cite{Battarra:2011nt,Amsel:2011km} $\tV = W_0 - W$ approaches
\begin{equation}
\tV = -f \beta^n-\frac{d \beta^2}{2 \log \beta^2}+\ldots
\end{equation}
at large $\beta$.  Once again this does not agree with the expression appearing on the right hand side of the energy bound \eqref{Ebound2}.

As before, one can instead define
\begin{equation}
\label{VeffDbf}
 \V(\beta) =  \alpha(\beta) \beta - W(\beta) +W_0(\alpha(\beta))  \,,
\end{equation}
which is simply the ``Legendre transform''  $W(\alpha) \to \alpha \beta - W(\beta)$ of the Neumann effective potential.  This function has the properties that its extrema (at $\beta = \beta_*$) correspond to solitons satisfying our boundary conditions and that $\V(\beta_*)$ gives the energy of the soliton.
 Furthermore, using the asymptotic scaling behavior (\ref{bfW0large}), we have in the planar limit
\begin{equation}
\label{veffBF}
 \V(\beta) = \alpha(\beta) \beta - W(\beta) +k_c\alpha(\beta)^2 +\frac{1}{d} \alpha(\beta)^2 \log|\alpha(\beta)| \,,
\end{equation}
which is just the expression that appears in the energy bound \eqref{bound0bf}.
It also follows  that the fake supergravity domain wall solutions saturate the energy bound.

The off-shell effective potential can be constructed with a similar gluing construction as in section~\ref{sec:gluing}.  Now the energy of the time- and rotationally-symmetric initial data is
\begin{eqnarray}
\frac{E_\alpha}{V_k} &=& \frac{d-1}{2}M_0 - \alpha \beta +\frac{d \beta^2}{4}+W(\alpha),\quad (\mathrm{Neumann}),\\
\frac{E_\beta}{V_k} &=& \frac{d-1}{2} M_0 +\frac{d \beta^2}{4}-W(\beta), \hspace{1.5cm} (\mathrm{Dirichlet}).
\end{eqnarray}
Matching solutions at large $r$ as before yields
\begin{equation}
\frac{E_\alpha}{V} = W(\alpha)+ W_0(\alpha) +\frac{(\beta_0(\alpha)-\beta)^2}{2 \log r_*}+\ldots
\end{equation}
for the deformed Neumann theory and
\begin{equation}
\frac{E_\beta}{V} = \alpha\beta - W(\beta)+
 W_0(\alpha(\beta))
 +\frac{(\beta_0(\alpha(\beta))-\beta)^2}{2 \log r_*}+\ldots \,.
 \end{equation}
for the deformed Dirichlet theory.

\end{document}